\documentclass[conference,a4paper]{IEEEtran}
\IEEEoverridecommandlockouts

\makeatletter
\def\markboth#1#2{\def\leftmark{\@IEEEcompsoconly{\sffamily}\MakeUppercase{\protect#1}}%
\def\rightmark{\@IEEEcompsoconly{\sffamily}\MakeUppercase{\protect#2}}}
\makeatother

\usepackage[english]{babel}
\usepackage{color}
\usepackage{lipsum}% http://ctan.org/pkg/lipsum
\usepackage{caption}
\usepackage{cite}
\usepackage[pdftex]{graphicx}
\usepackage{subcaption}
\usepackage{amsmath}
\usepackage{amsfonts}
\usepackage{array}
\usepackage{verbatim}
\usepackage{listings}
\usepackage{algorithm}
\usepackage{algpseudocode}
\usepackage{url}
\usepackage{enumerate}
\usepackage{multirow}
\usepackage{mathtools}

\usepackage{epsfig}
\usepackage{epstopdf}
\usepackage{multicol}% http://ctan.org/pkg/multicols
\usepackage[font=footnotesize]{caption}

\renewcommand{\arraystretch}{2}

\newcommand{\Hb}{\mathbf{H}}

\newcommand{\uu}{\mathbf{u}}

\newcommand{\w}{\mathbf{w}}

\newcommand{\bi}{\begin{itemize}}
\newcommand{\ei}{\end{itemize}}
\newcommand{\be}{\begin{equation}}
\newcommand{\ee}{\end{equation}}

\def\Exp{\mathbb{E}}

\def\beq{\begin{equation}}
\def\eeq{\end{equation}}
\def\beqa{\begin{eqnarray}}
\def\eeqa{\end{eqnarray}}
\def\beqan{\begin{eqnarray*}}
\def\eeqan{\end{eqnarray*}}

\usepackage{threeparttable}
\usepackage[table,xcdraw]{xcolor}
\usepackage{tabularx}
   \usepackage{multirow}
   \usepackage{booktabs}
   \usepackage{graphicx}
   \usepackage{array, blindtext}
   \usepackage{wrapfig}
\usepackage{pdfpages}

\title{Channel Dynamics and SNR Tracking in \\Millimeter Wave Cellular Systems}
\author{{{\bf Marco Giordani}$^\dagger$, {\bf Marco Mezzavilla}$^\diamond$, {\bf Aditya Dhananjay}$^{*\diamond}$, {\bf Sundeep Rangan}$^\diamond$, {\bf Michele Zorzi}$^\dagger$ }\\
$^\dagger$ University of Padova, Italy \quad $^\diamond$NYU Wireless, Brooklyn, NY, USA \quad $^*$MilliLabs Inc., NY, USA\\
emails: \small{$\{$\texttt{giordani}, \texttt{zorzi}$\}$\texttt{@dei.unipd.it}, $\{$\texttt{mezzavilla}, \texttt{srangan}$\}$\texttt{@nyu.edu}, \texttt{aditya@courant.nyu.edu}
}

}

\begin{document}
\maketitle

% As a general rule, do not put math, special symbols or citations
% in the abstract or keywords.
\begin{abstract}
The millimeter wave (mmWave) frequencies are likely to play a significant
role in fifth-generation (5G) cellular systems.
A key challenge in developing systems in these bands is the
potential for rapid
channel dynamics:  since mmWave signals are blocked by many
materials, small changes in the position or orientation of the
handset relative to objects in the environment can cause large
swings in the channel quality.  This paper addresses the issue
of tracking the signal to noise ratio (SNR), which is an essential
procedure for rate
prediction, handover and radio link failure detection.
A simple method for estimating the SNR from periodic
synchronization signals is considered.  The method is then evaluated
using real experiments in common blockage scenarios combined with outdoor
statistical models.
\end{abstract}

% Note that keywords are not normally used for peerreview papers.

% For peer review papers, you can put extra information on the cover
% page as needed:
% \ifCLASSOPTIONpeerreview
% \begin{center} \bfseries EDICS Category: 3-BBND \end{center}
% \fi
%
% For peerreview papers, this IEEEtran command inserts a page break and
% creates the second title. It will be ignored for other modes.
\IEEEpeerreviewmaketitle

\section{Introduction}

Each generation of wireless mobile technology has been
driven by the need to meet new requirements
that could not be completely achieved by its predecessor.
Following this trend, fifth generation cellular
(5G) systems are now expected
to meet unprecedented speeds, near-wireline latencies and ubiquitous connectivity with uniform user Quality of Experience (QoE) \cite{samsung,osseiran2014scenarios}.
While current microwave bands below 3~GHz
have become nearly fully occupied, the millimeter wave
(mmWave) frequencies, roughly above $10$ GHz, have
enormous amounts of unused available spectrum.
These bands are widely expected to become a key means of addressing the challenge of higher required data rates \cite{mmW,rappaportmillimeter,RanRapE:14}.

However, one of the key challenges for cellular
systems in the mmWave bands is the rapid channel dynamics.
In addition to the high Doppler shift,
mmWave signals are completely blocked by many common
building materials such as brick and
mortar~\cite{rappaport2014millimeter}, and even the human body can cause
up to 35~dB of attenuation~\cite{lu2012modeling}.
As a result, the movement of obstacles and reflectors,
or even changes in the orientation of a handset relative to a body or a hand, can cause the channel to rapidly appear or disappear.

This high level of channel variability has widespread implications
for virtually every aspect of cellular design.  This paper
focuses on one particular important design issue
which is the tracking of the
downlink channel quality and signal-to-noise ratio (SNR) at the mobile
user equipment (UE).
Measuring the SNR and reporting the value in periodic \emph{channel quality indicator} (CQI) reports is an essential component of any modern
cellular system -- see, for example \cite{LTE_book,schwarz2010calculation}
for a detailed description of the methods in 3GPP LTE.
Most importantly, CQI reports are the basis for rate prediction
and adaptive modulation and coding.
While CQI errors can be mitigated somewhat
via Hybrid ARQ (HARQ), HARQ requires retransmissions that may
result in excess delay.  One of the goals of 5G systems is to achieve
very low ($< 1$ ms) air link latencies.
CQI and related signals measurements are
also necessary for proper handover determination  and radio link failure detection, which are likely to become more common in mmWave
due to the small cell topology and the intermittency of the channel.

While CQI estimation is relatively straightforward in current
cellular systems, there are at least three
potentially complicating issues for mmWave:
(i) the rapid dynamics due to blockage events that strongly affect the link quality;
(ii) the need to track the CQI in multiple spatial directions with very
narrow beams; and
(iii) the limited number of available measurements
since the
\emph{cell reference signal} (CRS) used in current 3GPP LTE
 systems may not be available for mmWave (see Section~\ref{sec:cqilte}).

To address these challenges, this paper presents two key contributions.
First, we propose a novel method for estimating the channel quality
using synchronization signals and directional scanning.
This signaling mechanism was also considered for initial access in
\cite{Barati,barati2015initial}.
We derive an unbiased estimate for the instantaneous
wideband SNR in a particular pointing direction.
The estimate can then be filtered over time to trade off noise
reduction and tracking speed.

Secondly, we evaluate the SNR tracking through real measurements
using a novel high-speed measurement system.
There are currently a large number of
measurements of mmWave outdoor channels and detailed statistical models
\cite{Mustafa,Samimi2015Prob,MacCartney2015Wideband,ns3_nokia}.
However, these measurements have been largely
performed in static locations with minimal local blockage.
The dynamics of the channel are not fully understood
-- see some initial work in \cite{eliasi2015stochastic,ferrante2015mm}.
In this work, we experimentally measure the dynamics of the channel
in various common blockage scenarios using a high-speed channel
sounder at 60~GHz.  We then
combine the measured channel traces with the
statistical models to evaluate the SNR tracking algorithms.
%
%This paper seeks to evaluate how fast the channel actually changes under realistic blocking events and if UEs can track those changes through simple digital signal processing. We first present a simple mechanism for estimating the SNR in a cellular system, combining a simulated statistical channel  with  local blocking dynamics measured experimentally.  Since the CRSs are not available in mmWave systems, we propose to  estimate the SNR from periodic synchronization signals. Moreover, we present different filtering algorithms to track such estimate, if affected by noise, to recover a trace as much aligned as possible to the actual channel conditions, for faithfully estimating the channel.
%
%In detail, in Section \ref{sec:system_model}, after describing the synchronization signals structure we use to estimate the SNR trace, as a replacement for the CRSs,  we present  the simulated statistical channel model and the filtering algorithms used to reduce the noise from the SNR estimate. In Section \ref{sec:testbed} we describe the experimental testbed that we use to measure the real-world dynamics of different local blocking events,  that will be modulated on top of the simulated statical channel. In Section \ref{sec:sis_par}, we derive some parameters that we will use in the SNR tracking simulations, while in Section \ref{sec:res} we show our major findings and results. Finally, in Section \ref{sec:concl} we summarize the key aspects of this paper, presenting our expected future works.

\section{System Model}
\label{sec:system_model}

\subsection{CQI Estimation in 3GPP LTE} \label{sec:cqilte}

CQI estimation of the downlink channel is relatively straightforward
in 3GPP LTE \cite{LTE_book,schwarz2010calculation}:
the downlink channel quality is measured from what is
called the cell reference signal (CRS).  This is a wideband signal
transmitted essentially continuously with one signal being sent
from each BS cell transmit antenna port.
Each UE in connected mode
monitors these signals to create a wideband channel estimate that can be used
both for demodulating
downlink transmissions and for estimating the channel quality.

However, in addition to the rapid variations of the channel,
there are two issues for CQI estimation in mmWave.
First, a CRS will likely not be available since downlink
transmissions at mmWave frequencies
will be directional and specific to the UE.  Demodulation
reference signals
will thus likely follow the format of
LTE's UE-specific reference signals, which are
transmitted in-band with the data.
Thus, there will likely be no reference signals that are broadcast to all UEs.
Secondly, mmWave UEs are likely to use analog beamforming, meaning that
the UE can only measure the channel quality in one direction at a time
\cite{sun2014mimo,heath2015overview}.

\begin{figure}[t!]
\centering
 \includegraphics[trim= 0cm 0cm 0cm 0cm , clip=true, width=0.8 \columnwidth]{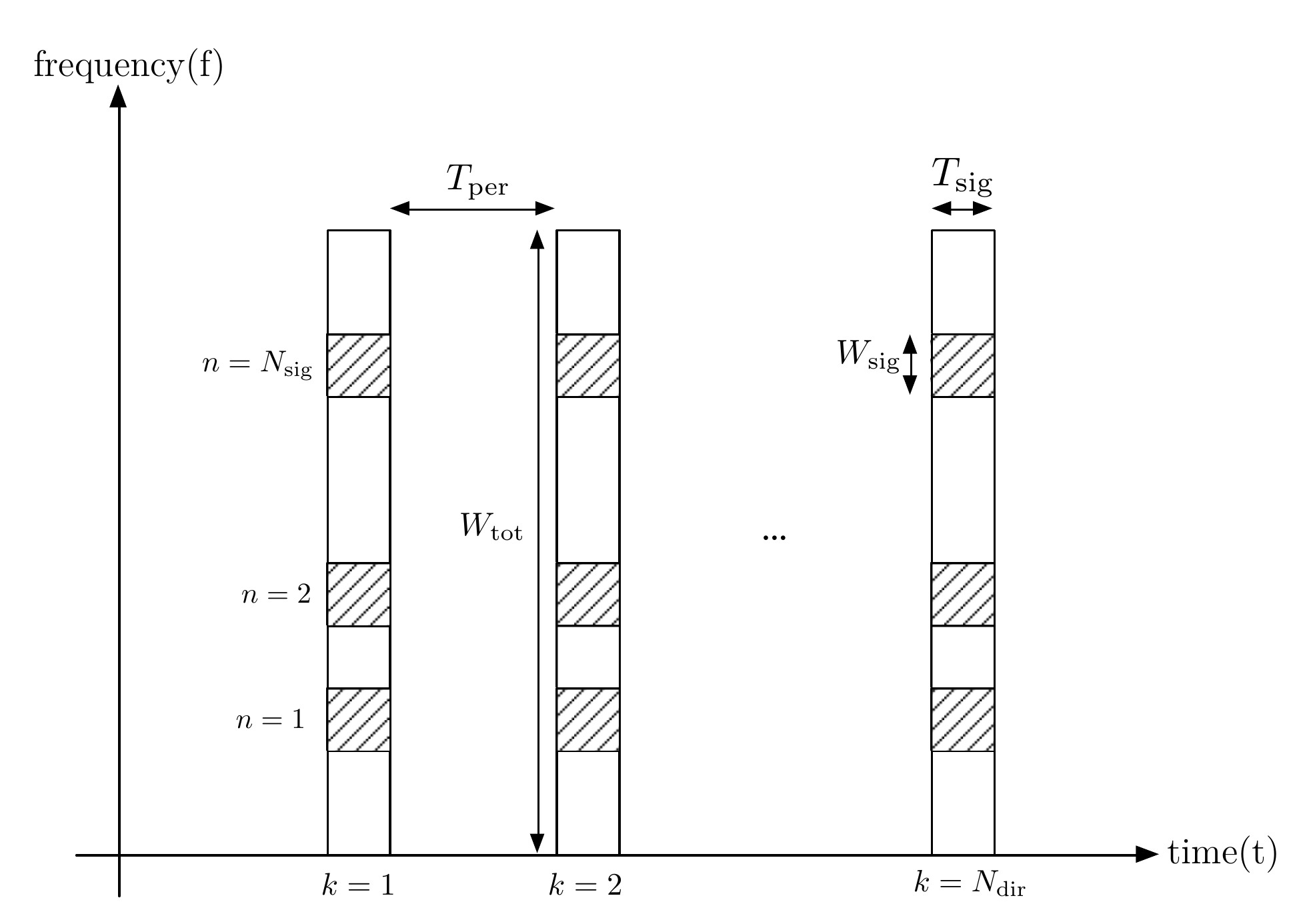}
 \caption{Periodic transmission of narrowband synchronization signals from the BS. This structure is similar to the LTE PSS.}
 \label{fig:slot}
\end{figure}

\subsection{Synchronization Signal Transmission Format}

In the absence of  CRS, each UE
must find an alternate signal to measure the downlink channel quality.
For this work, we propose that
the UE  estimates the channel quality from
periodic synchronization signals similar to the LTE primary or secondary
synchronization signals (PSS or SSS) used for initial access
and cell search.
These signals are transmitted at a much lower duty cycle and
the estimation of the channel
from these limited measurements
is one of the key challenges addressed in this paper.

For the structure of the synchronization signals, we assume the format
described in \cite{Barati} and reported in Figure \ref{fig:slot}.  Similar to the LTE PSS, we assume that
each BS cell transmits a synchronization signal once every $T_{\rm per}$ seconds
for a duration of $T_{\rm sig}$ seconds.
These signals will be transmitted
omni-directionally or in a fixed pattern covering the cell area.
Each transmission consists of $N_{\rm sig}$ sub-signals where each
sub-signal is transmitted over a narrow band of $W_{\rm sig}$ Hz.
The use of  multiple transmissions is for frequency diversity.

At the UE side, we assume that the UE
receiver attempts to estimate the received SNR of the
synchronization signals in $N_{\rm dir}$ different angular
directions.
As discussed above, we assume the UE  performs analog beamforming
and hence can measure the synchronization signal in only
one direction at a time.
We thus assume that in each synchronization signal
period, the UE measures the received signal strength in
one of the $N_{\rm dir}$ angular directions.  Hence, it can make
a received signal measurement in a particular angular direction once every
$N_{\rm dir}T_{\rm per}$ seconds.
The specific parameter values will be discussed in
Section \ref{sec:sis_par}.

\subsection{Channel Model and SNR Tracking}
\label{sec:SNR_tracking}

Let $p_{ik}(t)$ be the $k$-th transmitted sub-signal in the $i$-the
synchronization period.  Let $t_i$ denote the time of the synchronization period
and $f_k$ the frequency location of the sub-signal within that period.
We assume that the sub-signal is received at the receiver as
\[
    r_{ik}(t) = \w_i^{rx*}\Hb(t_i,f_k)\w_i^{tx} p_{ik}(t) + n_{ik}(t),
\]
where $\w_i^{rx}$ is the RX beamforming vector at the UE,
$\w_i^{tx}$ is the TX beamforming vector at the BS cell,
$\Hb(t_i,f_k)$ is the narrowband channel response for the synchronization signal, 
and $n_{ik}(t)$ is AWGN.  Note that, as described above, we assume that 
each sub-signal is transmitted in a sufficiently narrow band that we can assume flat fading  across the transmission.  We let $N_0$ denote the noise
power spectral density.

We assume a standard multi-path channel model \cite{Mustafa}
where the time-varying channel response is given by
\beq \label{eq:Hmp}
    \Hb(t,f) = \frac{1}{\sqrt{L}} \sum_{\ell =1}^L \sqrt{g_\ell(t)}
        e^{2\pi j (f_{d,\ell} t - \tau_\ell f)}
        \uu^{rx}_\ell \uu^{tx *}_\ell,
\eeq
where $L$ is the number of paths and, for each path $\ell$,
$g_\ell(t)$ is the time-varying channel power,
$f_{d,\ell}$ is the path Doppler shift, and $\uu^{rx}_\ell$ and $\uu^{tx}_\ell$
are the RX and TX spatial signatures of the path that  depend on the
angles of arrival and departure of the path from the antenna arrays.

In this work, we are interested in tracking the SNR in a single
TX and RX pointing direction.
As described in the previous
subsection, the BS cell will use a fixed transmit direction
and the UE receiver will scan  $N_{\rm dir}$ beamforming 
directions and estimate the SNR separately in each direction.
For the remainder of this paper, we focus on a subset of the
transmission times $i$ where
the TX and the RX are pointed in a particular direction  $\w_i^{tx} = \w^{tx}$
and $\w_i^{rx} = \w^{rx}$, for some fixed $\w^{tx}$ and $\w^{rx}$.

Given TX and RX directions $\w^{tx}$ and $\w^{rx}$,
define the wideband average channel gain as
\[
    G(t) = \frac{1}{W_{\rm tot}}
        \int_{f_c-W_{\rm tot}/2}^{f_c+W_{\rm tot}/2}
        |\w^{rx*}\Hb(t,f) \w^{tx}|^2df,
\]
where the integral is over the total
system bandwidth of $W_{\rm tot}$ at center frequency $f_c$.
If the base station transmits at a power $P_{tx}$, then the average wideband SNR would be
\beq \label{eq:gamwide}
    \gamma(t) := \frac{G(t)P_{tx}}{N_0W_{\rm tot}},
\eeq
where $W_{\rm tot}$ is the total system bandwidth.
We call $\gamma(t)$ the \emph{true wideband} SNR.

As stated before, since mmWave cells will not transmit a CRS,
we  wish to  estimate the wideband SNR $\gamma(t)$
from the synchronization signal.
The wideband SNR can be estimated as follows:
let $E_s = \int |p_{ik}(t)|^2dt$ denote the transmitted signal energy per sub-signal.
We assume this does not vary with $i$ or $k$.
If the transmit power is $P_{tx}$, the signal duration is $T_{\rm sig}$ and there are
$N_{\rm sig}$ signals,
\beq \label{eq:Es}
    E_s = \frac{P_{tx}T_{\rm sig}}{N_{\rm sig}}.
\eeq
Now suppose that the receiver applies
a matched filter for each sub-signal to obtain the statistic
\begin{align}
    \MoveEqLeft z_{ik} = \frac{1}{\sqrt{E_s}}\int p_{ik}^*(t)r_{ik}(t)dt      \label{eq:zik} \\
    &=   \sqrt{E_s}\w^{rx*}\Hb(t_i,f_k)\w^{tx}  + v_{ik},
    \quad  v_{ik} \sim {\mathcal CN}(0,N_0). \nonumber
\end{align}
It is easy to verify that if the frequency $f_k$ is uniformly randomly distributed
over the system bandwidth, then
\[
    \Exp\left[|z_{ik}|^2\right] = \frac{G(t)P_{tx}}{N_{\rm sig}} + N_0.
\]
Hence, we can form an unbiased estimate of $\gamma(t)$ in \eqref{eq:gamwide} by
\beq \label{eq:gamhati}
    \hat{\gamma}_i = \frac{1}{N_0T_{\rm sig}W_{\rm tot}}
        \sum_{k=1}^{N_{\rm sig}} \left[ |z_{ik}|^2 - N_0 \right],
\eeq
which sums the received power on the $N_{\rm sig}$ sub-signals and subtracts the noise.

\subsection{Filtering Algorithms}
\label{sec:filtering_algh}

\begin{table}[!t]
\centering
\renewcommand{\arraystretch}{1.2}% Tighter
\begin{tabular}{|l|l|}
\toprule
\textbf{Symbol} & \textbf{Description} \\
\hline
$\gamma_i$ & \begin{tabular}{@{}l @{}} Wideband true SNR \end{tabular} \\
\hline
$\hat{\gamma}_i$ & \begin{tabular}{@{}l @{}} Raw SNR estimate of  $\gamma_i$ \\ from the synchronization signals  \end{tabular} \\
\hline
$\bar{\gamma}_i$ & \begin{tabular}{@{}l @{}} Time-filtered SNR  estimate of $\gamma_i$ from $\hat{\gamma}_i$ \end{tabular} \\
\bottomrule
\end{tabular}
\caption{Symbols for the SNR and its estimates.}
\label{tab:SNR_traces}
\end{table}

Since $\hat{\gamma}_i$ in \eqref{eq:gamhati} is an estimate of the wideband SNR that has been obtained starting from the synchronization signals, it may deviate from the true SNR due to noise.
We call the measurement $\hat{\gamma}_i$ the \emph{raw} SNR.
To reduce the noise, we can filter the raw SNR producing a time-averaged value
that we will denote by $\bar{\gamma}_i$.  We consider three possible filtering schemes \cite{Dcom}:
\begin{itemize}
\item \emph{No filtering:}  In this case, we simply take $\bar{\gamma}_i = \hat{\gamma}_i$.
\item \emph{First-order filtering:}  This uses a simple low-pass filter:
\beq \label{eq:gamFO}
    \bar{\gamma}_i = (1-\alpha) \bar{\gamma}_{i-1} + \alpha \hat{\gamma}_i,
\eeq
for some constant $\alpha \in (0,1)$.
\item \emph{Moving average filtering:}  In this algorithm, we simply average the last $M$
values,
\beq \label{eq:gamMA}
    \bar{\gamma}_i = \frac{1}{M} \sum_{j=1}^M \hat{\gamma}_{i-j+1}.
\eeq
\end{itemize}

Therefore $\bar{\gamma}_i $ is a \emph{filtered SNR estimate} of $\gamma_i$, obtained starting from the noisy raw SNR $\hat{\gamma}_i$. Our goal is to find  the optimum scheme to  minimize the \emph{average estimation error} ${e_i =\Exp\left[|\bar{\gamma}_i -\gamma_i|\right]}$, in order to derive an SNR stream that can be used to reliably estimate the channel quality.

\section{Experimental Evaluation}
\label{sec:testbed}

\subsection{Channel Modeling Overview}

\begin{figure*}[t!]
\centering
 \includegraphics[trim= 0cm 0cm 0cm 0cm , clip=true, width= \textwidth]{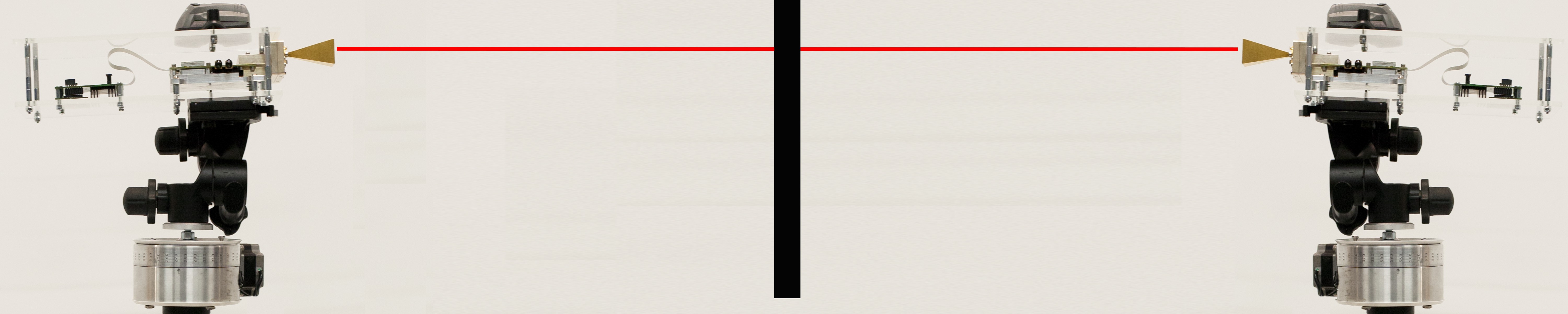}
 \caption{Our mmWave testbed. We introduce an obstacle (person walking, hand, metal plate) in front of the receiver to observe the received power drop.}
 \label{fig:TX}
\end{figure*}

While there has been considerable progress in understanding the mmWave channel
for long-range outdoor cellular links, most of the studies have been performed
in stationary locations with minimal local blockage.  For example, 
in the New York City studies in \cite{Mustafa,Samimi2015Prob,MacCartney2015Wideband,ns3_nokia},
the RX was placed in a fixed location on a cart.  In addition, there were
no obstacles in the immediate vicinity of the RX, such as a hand or a person,
whose movement would cause signal variations due to blockage.

Unfortunately, measuring a wideband spatial channel model with dynamics is not possible
with our  current experimental equipment.  Such a measurement would require that the
TX and RX directions   be swept rapidly during the local blockage event.  Since our
platform relies on horn antennas mounted on mechanically rotating gimbals, such
rapid sweeping is not possible.

In this work, we thus propose the following alternate approximate method to generate
realistic dynamic models for link evaluation:
\begin{enumerate}
\item We first randomly generate the number of paths, relative power, delay and
angles of arrival and departure based on the wideband channel models in \cite{Mustafa} and \cite{ns3_nokia}.
These models are based on extensive measurements in New York City in links similar to a
likely urban micro-cellular deployment, and would reflect the characteristics of a
stationary ground-level mobile with no motion nor local obstacles.

\item Combining the angles of arrivals and departure with the
antenna array patterns at the BS and UE, we can then determine the spatial signatures
$\uu_\ell^{rx}$ and $\uu_\ell^{tx}$ in \eqref{eq:Hmp}.  The randomly generated parameters
from Step 1 will also provide the delay  and  power of each path,
that we will denote by $\tau_\ell$ and $P_\ell$, respectively.

\item We assume a random direction of motion of the UE receiver.  Based on the UE velocity
and angle of motion relative to the angle of arrivals of the path, we can compute the Doppler
shifts $f_{d,\ell}$ in \eqref{eq:Hmp} by $f_{d,\ell} = f_{d,max}\cos\theta_\ell$, where $f_{d,max}$ is the maximum
Doppler shift and $\theta_\ell$ is the angle between the path
angle of arrival and direction of motion.

\item Finally, if there were no local blockage,
then the path powers $g_\ell(t)$ in \eqref{eq:Hmp} could be fixed as
$g_\ell(t) = P_\ell$, where the values
$P_\ell$ are the path powers generated in the static model in Step 1.
To simulate local blockage, we assume that these powers will be modulated as
\beq \label{eq:gell}
    g_\ell(t) = \beta P_\ell h(t),
\eeq
where $h(t)$ is a time-varying scaling factor accounting for the blockage and $\beta$ is a scaling factor.
Since there are no statistical models for the blockage dynamics, we measure
traces of $h(t)$ experimentally in various blockage scenarios.
The factor $\beta$ can then be adjusted to set a desired test SNR, according to the envisioned target rate a mmWave user is expected to reach. We refer to Section \ref{sec:sis_par} for further details on the choice of this parameter.
\end{enumerate}

This four step procedure thus provides a semi-statistical model, in which
(i) the spatial characteristics of the channel are determined from
static statistical models derived from outdoor measurements and
(ii) local blockage events are measured experimentally and modulated on top of the
static parameters.

An important simplification in  \eqref{eq:gell} is that we assume that
the local blockage $h(t)$ equally attenuates all paths,  which  may not be realistic.
For example, a hand may block only paths in a limited range of directions.
However, this work considers the SNR tracking in only one direction at a time.
In any fixed direction, most of the power is contributed only from paths within
a relatively narrow beamwidth and thus the approximation that the
paths are attenuated together may be reasonable.

\subsection{Channel Sounding System}

We will call the scaling term $h(t)$ in \eqref{eq:gell} the \emph{local blockage factor}\footnote{Note that the absolute value of $h(t)$ is  immaterial, since the total channel power will be scaled by the factor $\beta$ in \eqref{eq:gell} to target a particular SNR.}. The key challenge in measuring the dynamics of local blockage is that we need
relatively fast measurements.  To perform these fast measurements, we used the
experimental channel sounding system in Figure \ref{fig:TX}: a high-bandwidth baseband processor,  built on a PXI  (a rugged PC-based platform for measurement and automation systems) from National Instruments, which engineers a real-world mmWave link. The transmitter and receiver operate in two separate boxes, each of which have the parts listed below:
\begin{itemize}
\item[(i)] an  8-slot chassis, capable of holding a variety of expansion cards.
\item[(ii)] a $1.73$  GHz quad-core PXIe controller that runs a realtime
operating system (RTOS) called PharLap, and communicates with the  computer used to run the experiments  through an Ethernet connection to coordinate the operation of each peripheral card in the chassis.
\item[(iii)]  two FPGA cards for the baseband signal processing.
\item[(iv)] a FlexRIO adapter module (FAM) card and a converter between
the baseband signal and an IF signal, which are connected to the antenna.
\item[(v)] mmW Converters, to convert the IF signal to  mmWave in the range of $57-63$ GHz. The IF signal is mixed with the output of a local oscillator (LO), filtered, amplified, and sent over a waveguide output. We use 23~dBi directional 
    horn antennas (manufactured by Sage millimeter) to interface with the waveguide. This converter works in tandem with a power supply and a controller card. An identical converter at the receiver performs the down-conversion from mmWave frequencies to IF.
\end{itemize}
%Our testbed also uses several other components such as a spectrum analyzer, a function generator, various horn and patch antennas, and SMA cables.

To sound the channel, we used a standard frequency-domain
method:  the \emph{transmitter} sent a continuous
repeating pattern created from an IFFT of a 128 point pseudo random
QPSK sequence.  We will call each group of 128 samples a symbol.
The sample rate is  130~MHz corresponding
to a symbol period of approximately 1~$\mu$s.  Note that this symbol period
is larger than the maximum delay spread.
The \emph{receiver} segments the received time domain sequence
into symbols, takes the FFT of each symbol and derotates it by the frequency-domain representation of the transmitted sequence.
Since the transmitted signal
is periodic, the derotated signal at the receiver will provide
an estimate of the frequency-domain response of the channel.
To reduce the effect of the noise, the sequence is averaged over 32 symbols, 
providing one averaged response every $32 \times 1 = 32$ ~$ \mu$s.
The averaged response is then converted to time-domain, to produce
the power delay profile (PDP) of the channel.

The phase noise at the receiver can be large (the manufacturer specification
is up to $-80$ dBc).  This is a common problem in many mmWave RF units.  
A characterization of this receiver
in \cite{aditya} found the maximum frequency deviation to be up 50 kHz,
which would be too large to leave uncompensated.
To compensate for the phase noise, in each
32 symbol measurement period the receiver derotated the signal by
9 frequency hypotheses spaced uniformly from $-50$ to 50 kHz,
and a potential PDP was generated from each of the 9 different
hypotheses.  The PDP with the maximum peak was then selected amongst the
9 hypotheses.  After this phase compensation, the received
symbols are sufficiently coherent over the 32 $\mu$s period needed for  a new averaged response to be provided.

\subsection{Measurement of Local Blockage}
\label{sec:meas_local_block}

\begin{figure}[t!]
\centering
 \includegraphics[trim= 0cm 0cm 0cm 0cm , clip=true, width= \columnwidth]{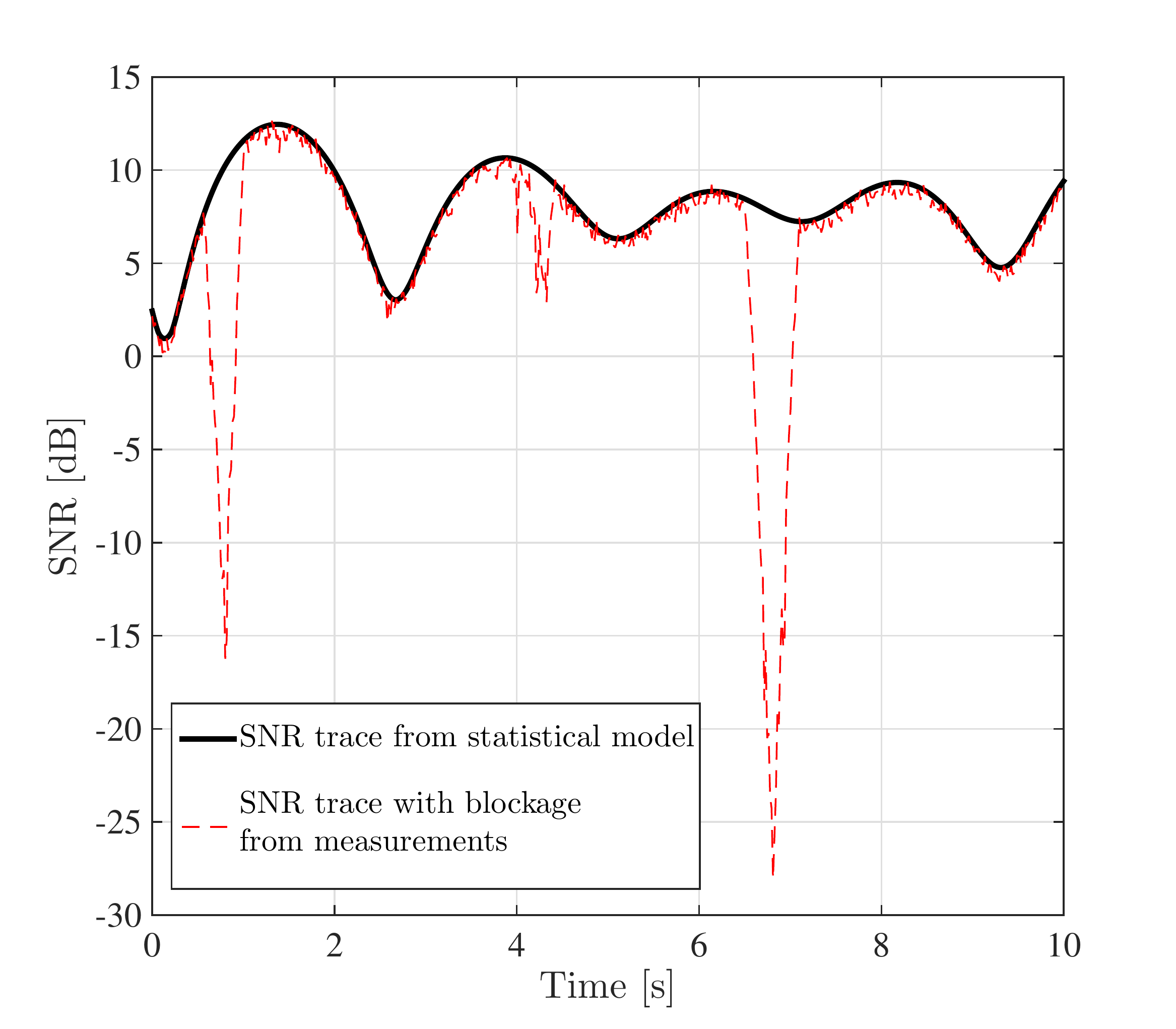}
 \caption{ SNR trace perceived when receiving the synchronization signals. The solid line is obtained by simulating the statistical channel described in \cite{Mustafa} and \cite{ns3_nokia} (without local obstacles). In the dashed line, the experimentally measured local blockage dynamics are modulated on  top of the statistical trace. The blockage is referred to a person walking multiple times between the transmitter and the receiver.}
 \label{fig:blockage}
\end{figure}

Using the above system, 
the blockage experiments were  conducted by
placing the transmitter and the receiver on a one-meter high pedestal, 
facing each other, at a distance of $4$ meters.
A laser pointer was  used to improve the alignment between the two devices.

After this set-up, the system is then run to continuously
collect PDPs during a blockage event.  
Blockage events are simulated by 
placing moving obstacles between the transmitter and the receiver. In 
this work, we considered three common blockage events: (i)  a person walking (or running) between TX and RX; (ii) a  wood (or metal) plate  held between the two communication edges; (iii) a hand holding a cellular phone.

The system was run during each of these blockage events
for a total time of 10 seconds.
During this time, PDPs were measured 
at a rate of one PDP per 32~$\mu$s.
We found that the dynamics of the channel varied considerably
slower than this rate, so we decimated the results 
by a factor of four, recording one PDP per 128 $\mu$s.
Since each experiment was run for $10$ seconds, 
 each experiment resulted in $10^7/128 = 78125$ PDP recordings.

To determine the local blockage function, we are only interested
in the line-of-sight path.  The power on this path was determined
from the maximum peak in the PDP.  Reflected paths would appear in other
samples and thus be rejected.  This received power then provides
the trace for the local blockage function $h(t)$ in \eqref{eq:gell}.
As described above, this local blockage function is then
used to modulate the time-varying channel response obtained
from the statistical channel model.

As an example, Figure \ref{fig:blockage} shows an SNR trace in which the blockage event is referred to a person walking multiple times between the transmitter and the receiver.

\subsection{Evaluation Results}
Once the SNR trace $\gamma(t)$ from the synchronization signals has been obtained, combining the simulated statistical channel  described in \cite{Mustafa} and \cite{ns3_nokia} with the local blockage dynamics measured experimentally, the raw SNR $\hat{\gamma}(t)$ can be estimated from the synchronization signals following Section \ref{sec:SNR_tracking}. In Section \ref{sec:sis_par} we describe the system parameters that we use in our simulations, while in Section \ref{sec:res} we analyze and compare the performance of the presented linear filters, applied to the  raw SNR trace, to obtain the estimate $\bar{\gamma}(t)$.

\section{Simulation parameters}
\label{sec:sis_par}

\begin{figure*}[t!]
\centering
 \includegraphics[trim= 0cm 0cm 0cm 0cm , clip=true, width=1 \textwidth]{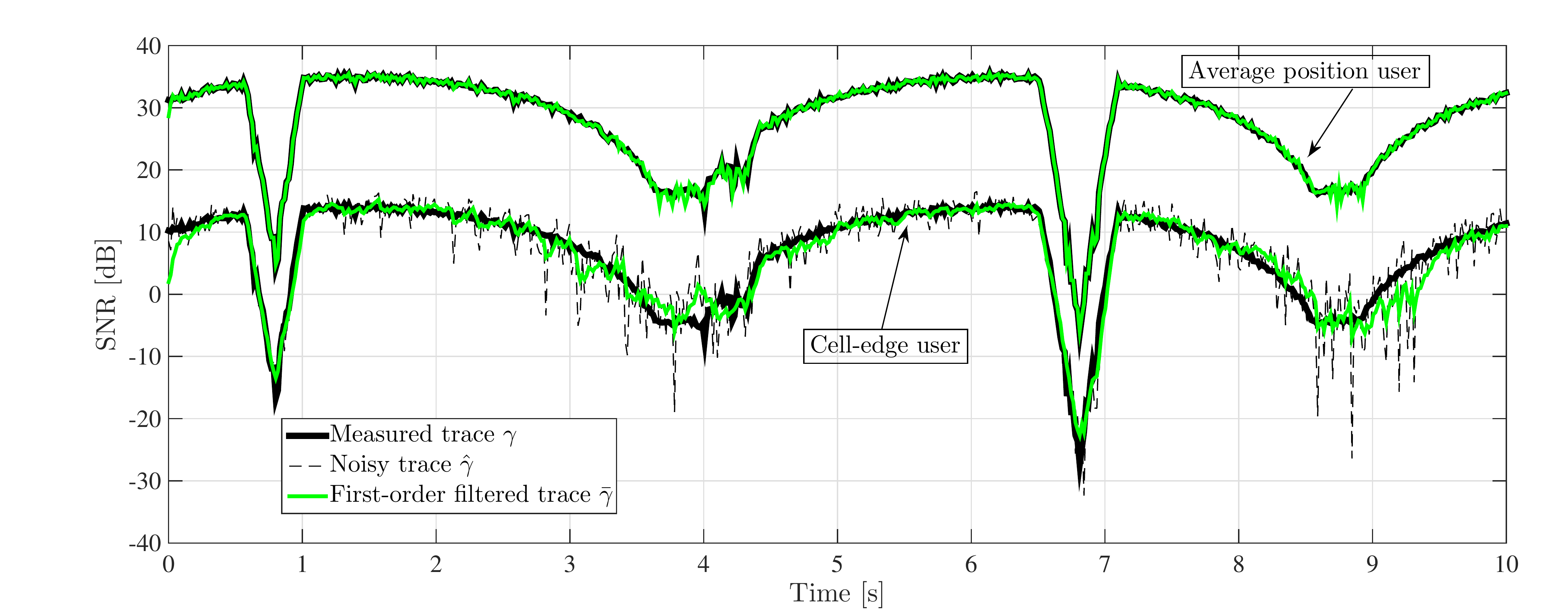}
 \caption{ SNR trace $\gamma(t)$ together with its raw version  $\hat{\gamma}(t)$ and its estimation $\bar{\gamma}(t)$ after different filtering schemes are applied. The upper lines refer to the SNR of a $50^{\text{th}}$ percentile typical user, while the lower lines refer to the SNR of a $5^{\text{th}}$ percentile edge user.}
 \label{fig:trace_filters}
\end{figure*}

In this section, we derive some parameters that we will use in the SNR tracking simulations: (i) the scaling factor $\beta$ of Equation \eqref{eq:gell}, to set a desired test SNR, and (ii) the SNR trace downsampling factor.

\subsection{SNR scaling factor $\beta$}

\begin{table}[!t]
\centering
\renewcommand{\arraystretch}{1.2}% Tighter
\begin{tabularx}{1\columnwidth}{@{\extracolsep{\fill}}ccc}
\toprule
\textbf{Description}& \textbf{$50^{\text{th}}$ percentile} & \textbf{$5^{\text{th}}$ percentile} \\
\toprule
\begin{tabular}{@{}c @{}} LTE spectral efficiency \\ $\rho$ (bit/s/Hz/$W_{\rm tot}$)  (from \cite{3GPP})  \end{tabular} & $\textbf{3.28}$ & $\textbf{0.154}$ \\
\midrule
\begin{tabular}{@{}c @{}} LTE rate (Mbps)\\  ($R_\mu = \rho \cdot W_{\rm tot} $) \end{tabular} & $3.28 \cdot 50 = \textbf{164}$ & $0.154 \cdot 50 = \textbf{7.7} $ \\
\midrule
\begin{tabular}{@{}c @{}} mmWave rate (Mbps)\\ (from \cite{Boccardi}, $R_{mmW} \simeq 9 R_\mu) $ \end{tabular} & $\textbf{1480}$ & $\textbf{70}$ \\
%\midrule
%\rowcolor[HTML]{EFEFEF} \begin{tabular}{@{}c @{}} mmWave target SNR  (dB)\\ (from Equation \eqref{eq:shannon}) \end{tabular} & $10 \log_{10}\Big(2^{\tfrac{1480}{0.8 \cdot 500 }}-1\Big) \simeq \: \textbf{10} $ & $ 10 \log_{10}\Big(2^{\tfrac{70}{0.8 \cdot 500 }}-1\Big) \simeq \textbf{-9}$ \\
\bottomrule
\end{tabularx}
\caption{Cell user 4G-LTE and expected 5G rate, for average-cell-position users ( $50^{\text{th}}$ percentile) and cell-edge users ($5^{\text{th}}$ percentile). For the LTE case, we refer to a DL SU-MIMO $4 \times 4$ TDD baseline for a microwave system using $50$ MHz of bandwidth. For the mmWave case, we refer to a system with $500$ MHz of bandwidth and a single user.}
\label{tab:rate}
\end{table}

As  previously asserted, the 
scaling factor $\beta$ in \eqref{eq:gell} 
is selected to bring the average SNR to some desired test level.
We consider two test cases:
\begin{enumerate}
\item the user belongs to the $50^{\text{th}}$ percentile, so it presents average propagation conditions;
\item the user belongs to the $5^{\text{th}}$ percentile, to simulate the $5\%$ worst user rate at cell edges.
\end{enumerate}

For each case, the target test SNR is obtained by: (i) determining a reliable 4G-LTE target rate for the user, according to \cite{3GPP}; (ii) determining the corresponding mmWave target rate, according to \cite{Boccardi}; and (iii) finding the corresponding target test SNR through a Shannon capacity evaluation.  

In order to find  a reliable data rate for the two user cases we are considering, we refer to the actual LTE 3GPP user data rate. We consider the performance of the DL SU-MIMO $4 \times 4$ TDD baseline in \cite{3GPP}, for a microwave system using $50$ MHz of bandwidth. In Table \ref{tab:rate}, we show the cell user spectral efficiency $\rho$ that we use to compute the LTE data rate $R_\rho$ (actually $R_\rho=\rho \cdot W_{\rm tot}$).

The corresponding DL data rate for a mmWave system with $500$ MHz of bandwidth can be determined referring to \cite{Boccardi}, where it is reported  that the data rate of a mmWave user is expected to be around $9$ times higher than the rate of current LTE systems.  According to this result,  Table \ref{tab:rate} reports  the corresponding rates that can realistically be achieved by mmWave users.

Based on the mmWave user data rate, we can estimate the corresponding target SNR $\gamma_t$ using the Shannon capacity:
\begin{equation}
R = \delta \cdot W_{\rm tot} \cdot \log_2(1+\gamma_t) \Longrightarrow \gamma_t = 2^{\tfrac{R}{\delta \cdot  W_{\rm tot} }}-1
\label{eq:shannon}
\end{equation}
where $R$ is the data rate, $W_{\rm tot}$ is the system 
bandwidth ($500$ MHz for the mmWave system we are considering), and $\gamma_t$ is the target data SNR. $\delta = 0.8$ is a parameter that accounts for a $20\%$ control overhead \cite{Mustafa}.
%The values of $\gamma_t$ assigned for the $50^{th}$ and $5^{th}$ percentiles respectively are shown in Table \ref{tab:rate}.
Solving \eqref{eq:shannon}, we obtain the target SNR $\gamma_t$.

The value $\gamma_t$  is the wideband SNR we would expect on a data channel.
The data channel would be received with the BS and UE performing beamforming.
However, the synchronization signals would be transmitted omni-directionally
and thus would be received at a lower SNR.  In the experiments below,
following \cite{Mustafa}, 
we will assume that the BS cell has $N_{tx}=64$ antennas, allowing
up to a 18 dB beamforming gain.  This gain would not be available for
the synchronization and thus the synchronization signals would be received
at a much lower SNR -- this is one of the main challenges in SNR tracking.
Thus, we assume that the wideband SNR (in linear scale) should be 
$\gamma_t/N_{tx}$.  

To set this SNR, we first 
generate a random trace of using the statistical model.  Then, we set the
factor $\beta$ in \eqref{eq:gell} to scale the average value of the
wideband SNR $\gamma(t)$ in the experiment to the desired target level $\gamma_t/N_{tx}$.
This generates the sequence for the wideband SNR $\gamma(t)$.
The raw estimate of the SNR $\hat{\gamma}(t)$ 
is then computed according to Section \ref{sec:system_model}.

%
%\subsection{$N_0$ Estimation}
% $N_0$ is computed according to  \cite{benvenuto_zorzi}:
%\begin{equation}
%N_0 = 10\log_{10}(k_B \cdot T \cdot BW)
%\end{equation}
%where $k_B$ is Boltzmann's constant, $T$ is the room temperature $(T = 300 K)$ and $BW=500$ MHz is the system bandwidth. Therefore, we get $N_0 \simeq  -87$ dBm.
%Following  \eqref{eq:zik} and \eqref{eq:gamhati}, for a specific user regime, we can estimate the raw SNR $\hat{\gamma}_i$ by generating random samples $v_{ik}$ of the AWGN noise with variance $N_0$.

\subsection{SNR Trace Downsampling Factor}
As we stated in Section \ref{sec:meas_local_block}, the measured SNR trace is composed of $78125$ samples, one every $128 \: \, \mu s$. According to Section \ref{sec:system_model} and the results in \cite{CISS}, we assume that each synchronization signal is transmitted periodically once every $T_{\rm per}=1$ ms for a duration of $T_{\rm sig} = 10 \:\mu$s, to maintain an overhead of $1\%$.  Moreover, we also assume that the user directionally receives such signals by performing an exhaustive search of the angular space through $N_{\rm slot}=16$ directions. Therefore, the transmitter and the receiver will be perfectly aligned just once every $N_{\rm slot} T_{\rm per} = 16 $ ms. For this reason, the original SNR trace has been downsampled, keeping just one sample every $16$ ms.

\section{Performance Evaluation}
\label{sec:res}

In Section \ref{sec:res_filter}, we analyze and compare the performance of the filtering algorithms described in Section \ref{sec:filtering_algh}. The goal is to determine the filter that minimizes the  estimation error ${e(t) =\Exp\left[|\bar{\gamma}(t) -\gamma(t)|\right]}$, for different user propagation characteristics ($50^{\text{th}}$ and $5^{\text{th}}$ percentiles). Furthermore, Section \ref{sec:res_error} shows how the estimation error changes when considering different SNR regimes.

\subsection{Filters performance comparison}
\label{sec:res_filter}

In Figure \ref{fig:trace_filters}, we plot the  SNR trace $\gamma(t)$, whose blockage events refer to a person walking multiple times between the transmitter and the receiver. The upper line refers to a $50^{\text{th}}$ percentile typical user  and the lower line to a $5^{\text{th}}$ percentile edge user. The figure also shows the noisy version $\hat{\gamma}(t)$ of the SNR trace, together with its estimate $\bar{\gamma}(t)$ after the presented linear filters have been applied. Two different scaling factors $\beta$ have been applied, when computing the SNR trace from the synchronization signals, according to the two user propagation regimes.

We see that, for low SNR regimes, the raw  SNR trace $\hat{\gamma}(t)$  shows a very noisy trend, which is considerably different from its original version. The reason is that, when the user receives the synchronization signals with very low power (e.g., when located at the cell edge), the noise component  dominates the SNR unbiased estimate in \eqref{eq:gamhati}, and therefore the raw SNR  substantially differs from $\gamma(t)$. A filtering algorithm is thus required, to recover a stream $\bar{\gamma}(t)$ that can be used to more accurately estimate the channel.
The main concern is that it is hard to discern between the downspikes which refer to an actual blockage  and those which accidentally manifested due to the additive noise. In such a way, the detection of real radio link failure situations might be distorted and might lead to false alarm or missed blockage detection events.

However,  a simple first-order filter can produce an estimated SNR trace $\bar{\gamma}(t)$ that appears very similar to the measured one. Therefore, even without designing much more complex and expensive nonlinear adaptive filters, we can properly restore the desired SNR stream and perform  reliable link failure detection and channel estimation. It should finally be noted that this filter requires a transient phase before reaching its normal operation, as can be seen in the first milliseconds of the traces in Figure \ref{fig:trace_filters}.

It is interesting to note that, when  considering a good SNR regime (upper lines of Figure \ref{fig:trace_filters}, simulating a $50^{\text{th}}$ percentile user), even the raw SNR  trace $\hat{\gamma}(t)$ (when no filters are applied) almost overlaps with its measured original version. Therefore, when an average position mmWave user receives uncorrupted synchronization signals, it estimates an SNR trace $\hat{\gamma}(t)$ that  sufficiently resembles the measured one, and can hence perform an adequately reliable channel estimation without any further signal processing.

\begin{figure}[t!]
\centering
 \includegraphics[trim= 0cm 0cm 0cm 0cm , clip=true, width=1 \columnwidth]{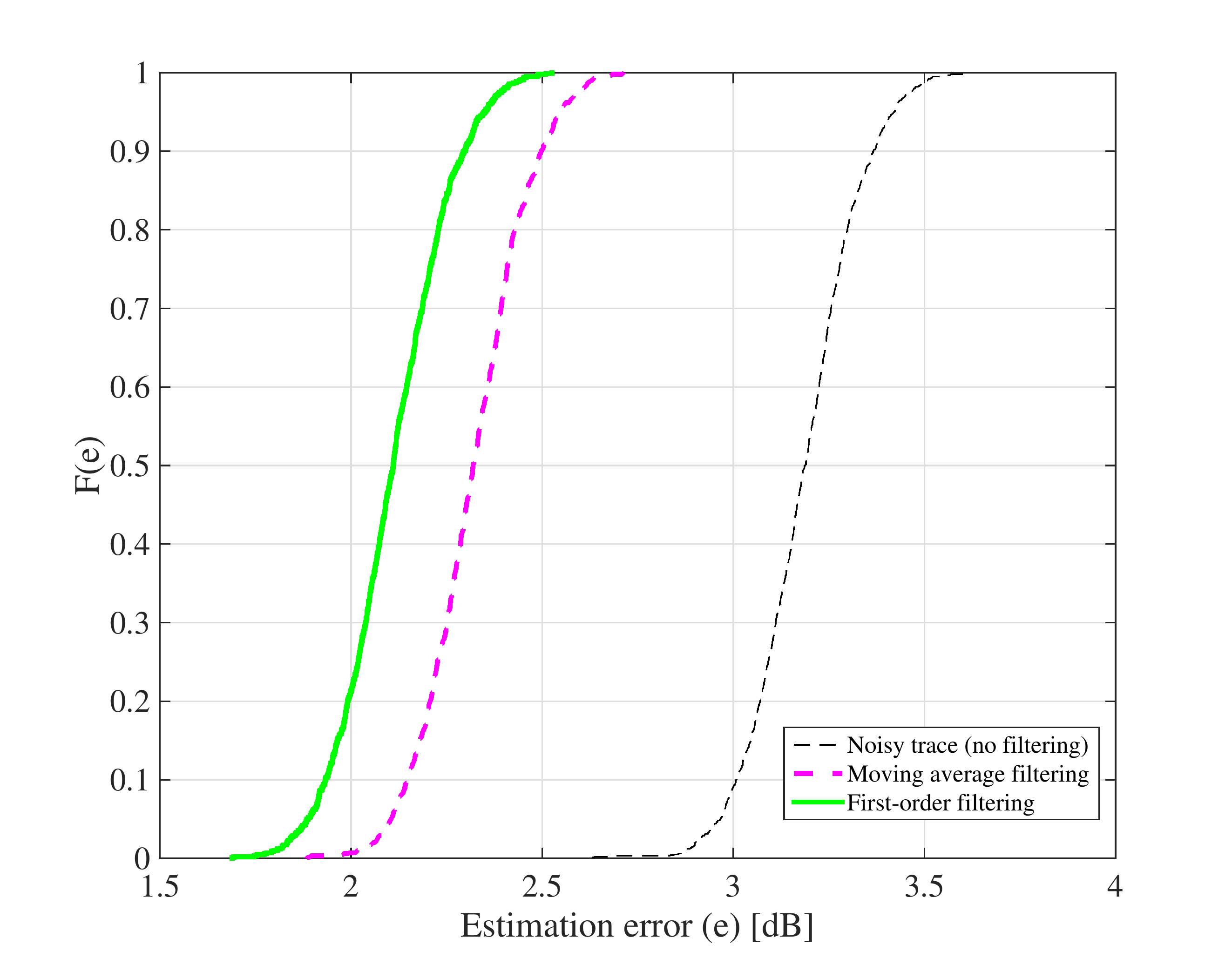}
 \caption{CDF of the estimation error ${e(t) =|\bar{\gamma}(t) -\gamma(t)|}$ for a $5^{\text{th}}$ percentile edge user, when different linear filters are applied to the noisy SNR trace $\hat{\gamma}(t)$.}
 \label{fig:CDF}
\end{figure}

In Figure \ref{fig:CDF} we plot the CDFs of the estimation error for a $5^{\text{th}}$ percentile edge user, when different linear filters are applied to the noisy SNR trace $\hat{\gamma}(t)$. The trace after a first-order filter is used shows much better performance with respect to the moving-average filtered trace which, besides its poor efficiency, is also affected by  a non-negligible  delay.
Therefore, among the options we considered, a first-order filter is the best choice to  reduce the estimation error and properly track the SNR trace.

\subsection{Analysis of the estimation error }
\label{sec:res_error}

\begin{figure}[t!]
\centering
\includegraphics[trim= 0cm 0cm 0cm 0cm , clip=true, width=1 \columnwidth]{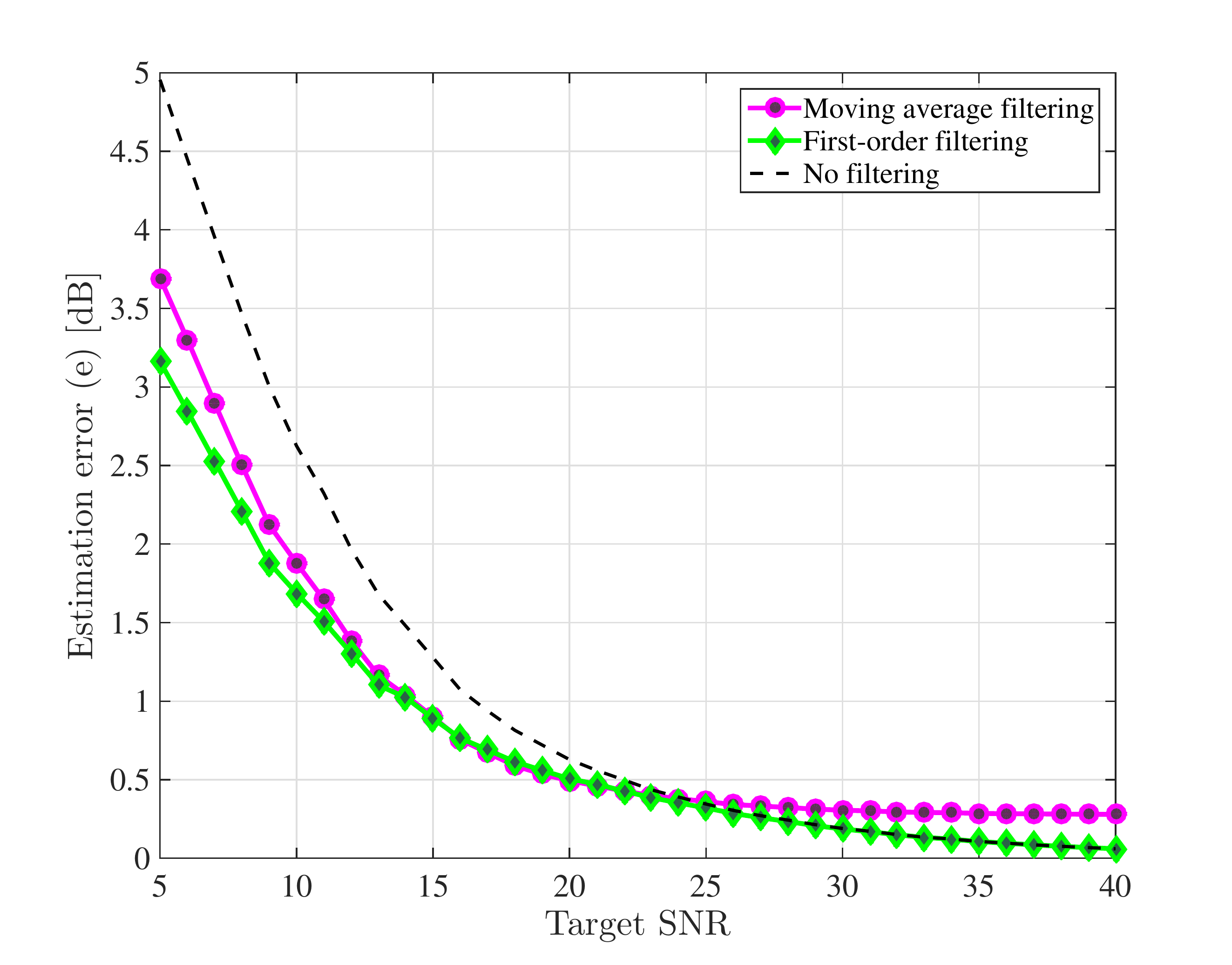}
 \caption{Average estimation error ${e(t) =\Exp\left[|\bar{\gamma}(t) -\gamma(t)|\right]}$ vs. target SNR, for different linear filter configurations.}
 \label{fig:error_vs_gamma}
\end{figure}

In Figure \ref{fig:error_vs_gamma}, we show the average estimation error ${e(t) =\Exp\left[|\bar{\gamma}(t) -\gamma(t)|\right]}$ versus different target SNR values $\gamma_t$, obtained by adjusting the scaling factor $\beta$ in Equation \eqref{eq:gell}. Multiple linear filter algorithms are applied to the raw SNR trace.

For low SNR regimes, we recognize again the better performance of the first-order filter, with respect to the capabilities of the moving-average filter. However, it is interesting to note that, after a certain threshold ($\gamma_t \geq 24$ dB), the moving-average is an even worse estimate than the noisy SNR trace $\hat{\gamma}(t)$ where no filters or further digital signal processing have been applied.
Moreover, in the same high-SNR range, the trend of $\hat{\gamma}(t)$  almost overlaps with the performance of the first-order filter in Figure \ref{fig:error_vs_gamma}. This agrees with the result of Subsection \ref{sec:res_filter} where we stated that, when simulating a $50^{\text{th}}$ percentile user, the  AWGN noise does not significantly affect  the estimated raw SNR trace $\hat{\gamma}(t)$, which therefore faithfully tracks the actual SNR evolution.

\section{Conclusions and Future Work}
\label{sec:concl}

A key concern for the feasibility of mmWave system is the
rapid channel dynamics.  Two broad questions need understanding:
how fast do channels actually change and how can systems be designed
to deal with these variations.  This paper has attempted to develop some fundamental  understanding in
the context of one particularly important problem -- namely
the tracking of SNR.  We have considered a simplified
procedure to estimate the SNR that can be readily implemented
in next generation systems using synchronization signals.
These signals will be necessary for initial access and thus will not
introduce further overhead.  Simple estimates for the SNR for these
were derived.  The methods were then evaluated in a novel 
semi-statistical model, where the spatial characteristics were derived from an 
existing statistical model based on 
outdoor measurements and the local blockage was derived from new 
experimental measurements.

Our high level finding is that the SNR can be mostly tracked within 
a few dB of error, even  when the measurements
are in very low SNR.  Nevertheless, using very simple filtering mechanisms,
the SNR tracking does incur some delay, particularly during periods of
very rapid changes.

Further work is still needed.  Most directly, it is useful to test nonlinear
and/or adaptive mechanisms that could track this SNR more effectively.
Also, this SNR tracking can then be used to assess the effects on other
higher layer functions including rate prediction, handover and
radio link failure detection.

\bibliographystyle{IEEEtran}
\bibliography{bibliography}

\end{document}